\documentclass[aip,preprint,preprintnumbers]{revtex4-1}
\usepackage{amsmath,amssymb,mathrsfs,braket}

\makeatletter
\newcommand{\dif}[1]{\frac{d}{d{#1}}}

\newcommand{\tr}[0]{\mbox{Tr}}
\newcommand{\ttr}[1]{\tr\left\{{#1}\right\}}
\newcommand{\Hilb}[0]{\mathscr{H}}
\newcommand{\bdop}[0]{\mathfrak{B}(\Hilb)}
\newcommand{\trcl}[0]{\mathfrak{T}_1(\Hilb)}
\newcommand{\hlsm}[0]{\mathfrak{T}_2(\Hilb)}
\newcommand{\rsdm}[0]{\mathfrak{D}(\Hilb)}

\newcommand{\ad}[0]{a^\dag}
\newcommand{\eq}[0]{e^{-\beta\omega\ad a}}
\newcommand{\eql}[1]{(\ad)^{#1}\eq}

\newcommand{\eqw}[2]{\eql{#1} a^{#2}}
\newcommand{\SK}[0]{\mathcal{K}}
\newcommand{\SL}[0]{\mathcal{L}}
\newcommand{\SO}[0]{\mathcal{O}}
\newcommand{\SX}[0]{\mathcal{X}}
\newcommand{\SY}[0]{\mathcal{Y}}
\newcommand{\Span}[1]{\mbox{Span}\left({#1}\right)}
\newcommand{\tpn}[0]{\frac{\gamma'}{\gamma}}
\newcommand{\tp}[0]{\left(\tpn\right)}
\newcommand{\tmpn}[0]{1-\frac{\gamma'}{\gamma}}
\newcommand{\tmp}[0]{\left(\tmpn\right)}
\newcommand{\gmdfn}[0]{\gamma-\gamma'}
\newcommand{\gmdf}[0]{(\gmdfn)}
\newcommand{\zeroN}[0]{\mathbb{Z}_{\geq 0}}
\newcommand{\bbra}[1]{\mathinner{\langle\langle{#1}|}}
\newcommand{\kket}[1]{\mathinner{|{#1}\rangle\rangle}}
\newcommand{\bbrakket}[1]{\mathinner{\langle\langle{#1}\rangle\rangle}}
\newcommand{\btra}[1]{\mathinner{\langle\langle\langle{#1}|}}
\newcommand{\ktet}[1]{\mathinner{|{#1}\rangle\rangle\rangle}}
\newcommand{\btraktet}[1]{\mathinner{\langle\langle\langle{#1}\rangle\rangle\rangle}}
\makeatother
\begin{document}
\preprint{UT-Komaba/10-2}
\preprint{WU-HEP-10-01}
\title{Spectral resolution of the Liouvillian of the Lindblad master equation\\for a harmonic oscillator}

\author{Daigo Honda}
%\email{}
\affiliation{Institute of Physics, University of Tokyo,\\ Komaba, Meguro-ku, Tokyo 153-8902, Japan}
  
\author{Hiromichi Nakazato}
 
\author{Motoyuki Yoshida}
\affiliation{Department of Physics, Waseda University,\\ Okubo, Shinjuku-ku, Tokyo 169-8555, Japan}
 
%\date{\today}

\begin{abstract}
A Lindblad master equation for a harmonic oscillator, which describes the dynamics of an open system, is formally solved.
The solution yields the spectral resolution of the Liouvillian, that is, all eigenvalues and eigenprojections are obtained.
This spectral resolution is discussed in depth in the context of the biorthogonal system and the rigged Hilbert space,
and the contribution of each eigenprojection to expectation values of physical quantities is revealed.
We also construct the ladder operators of the Liouvillian, which clarify the structure of the spectral resolution.
\end{abstract}

\maketitle

%
%
%%%%%%%%%%%%%%%%%%%%%%%%%%%%%%%%%%%%%%%%%%%%%%%%%%%%%%%
\section{Introduction}
%%%\label{***_Int}
%%%%%%%%%%%%%%%%%%%%%%%%%%%%%%%%%%%%%%%%%%%%%%%%%%%%%%%
%
%
The time evolution of an open quantum system interacting with external quantum systems is non-unitary. Therefore we cannot apply directly the usual
quantum mechanics to the open systems. Instead, we think the total system which is the joint system of a target system and an environment to be
a closed system to which the quantum mechanics is applicable, and extract the time evolution of the target system by taking the partial trace over the environment.
It is called the Reduced dynamics. Kraus \cite{Krau1} showed that if there is no initial correlation the time evolution of 
a reduced density matrix is complete positive. Lindblad \cite{Lind1} and Gorini et al.\ \cite{Gori1} obtained the general form of 
the generator of the complete positive dynamical semigroup (Liouvillian) $\SL$, which describes the time evolution of density matrices. 
A master equation with this generator is called a Lindblad master equation.

In this work, a system of a harmonic oscillator is considered, which interact linearly with a heat bath of bosonic field, an environment.
The Markov approximation and the van Hove limit yield the Lindblad master equation \cite{Facc1,Lax1,Loui1,Weid1} of the form
\begin{equation}
\begin{aligned}
&\dot{\rho}=\SL\rho\equiv-i[H,\rho]+\gamma\left(a\rho \ad-\frac{1}{2}\{\ad a, \rho\}\right)+\gamma'\left(\ad\rho a-\frac{1}{2}\{a \ad, \rho\}\right)\\
&H=\omega \ad a,\ \frac{\gamma'}{\gamma}=e^{-\beta\omega}.
\end{aligned}\label{lind_Int}
\end{equation}
Here, let $\Hilb$ be the separable Hilbert space of the states in the target system. 
$H$ is the Hamiltonian of the system, $\omega\in\mathbb{R}_{>0}$ a angular frequency, and $\ad, a$ are respectively the creation and annihilation operators,
which are densely defined unbounded operators on $\Hilb$ and satisfy the commutation relation $[a,\ad]=1$.
We can choose a set of the eigenstates of the Hamiltonian $\{\ket{n}\}_{n\in\zeroN},\ H\ket{n}=\omega n\ket{n},\ \ad\ket{n}=\sqrt{n+1}\ket{n+1},\ a\ket{n}=\sqrt{n}\ket{n-1}$
as a complete orthonormal system (C.O.N.S.) of $\Hilb$.
$\gamma, \gamma'\in\mathbb{R}_{> 0}$ are related to the damping rate and $\beta\in\mathbb{R}_{> 0}$ is an inverse temperature. A density matrix (state) is a trace class operator on $\Hilb$,
i.e. $\rho\in\trcl$, which satisfies $\tr\rho=1,\ \bra{\psi}\rho\ket{\psi}\geq0,\ \forall\ket{\psi}\in\Hilb$. Therefore (\ref{lind_Int}) is a differential equation on $\trcl$.
But we will restrict the space of density matrices later, because we treat the unbounded operators as observables.

This equation is well-known, and its derivation can be found in, e.g., Ref. \cite{Breu1}.
As an application of this equation, for example, the analysis of the cavity quantum electrodynamics may be considered.
The goal of this work is to solve this equation in a way to clarify the structure of the Liouvillian $\SL$.

There have been trials to solve this equation (see Refs. \cite{Brie1, Chat1, Endo1, Tay1}). Briegel and Englert have obtained the eigenvalues and the 
biorthogonal eigenbasis in terms of creation and annihilation operators by using normal ordering\cite{Brie1}, and Tay and Petrosky also have 
done in terms of the Wigner function\cite{Tay1}. But in this work we derive the solution by eigenprojection method in terms of creation and
annihilation operators without normal ordering, and present the spectral resolution of the Liouvillian $\SL$. 
If $\SL$ has an spectral resolution
\begin{equation}
\SL=\sum_m \lambda_m\Pi_m,\ \ \SL\Pi_m=\Pi_m\SL=\lambda_m\Pi_m,\ \Pi_m\Pi_n=\delta_{m,n}\Pi_m,\ \sum_m\Pi_m=\mathbf{1}, \label{spec_Int}
\end{equation}
where $\lambda_m$ is $m$-th eigenvalue and $\Pi_m$ is corresponding eigenprojection, 
the solution of the master equation is
\begin{equation}
\rho(t)=e^{t\SL}\rho(0)=\sum_me^{\lambda_m t}\Pi_m\rho(0) \label{sol_Int}.
\end{equation}

The paper is organized as follows.
In section \ref{SEC_Der} we derive the solution of (\ref{lind_Int}) in the form of (\ref{sol_Int}), and obtain the presumed eigenvalues and eigenprojections. 
In section \ref{SEC_PrEi} we confirm that they satisfy the property of the eigenprojections of the Liouvillian (\ref{spec_Int}). These eigenprojections have the 
structure of the biorthogonal system. Although the right eigenvectors are in $\trcl$, the left eigenvectors are unbounded operators. This is due to the unbounded Hamiltonian. 
In section \ref{SEC_Math} we give a mathematical justification of this spectral resolution on the basis of the rigged Hilbert space (Gel'fand triplet).
In section \ref{SEC_Prop} we discuss about the time behavior of this solution.
In section \ref{SEC_Calc} we calculate the time behavior of the expectation values and variances of some physical quantities exactly.
Although the number of eigenprojections is infinite, we need only finite number of eigenprojections.
Finally, in section \ref{SEC_Lad} we show the ladder operators of the Liouvillian. The eigenvalues of the Liouvillian have two indices, and there are
two sets of ladder operators.
The mathematical notions used in this paper are summarized in Appendices.
%
%
%%%%%%%%%%%%%%%%%%%%%%%%%%%%%%%%%%%%%%%%%%%%%%%%%%%%%%%
\section{Derivation of a formal solution and eigenprojections}\label{SEC_Der}
%%%\label{***_Der}
%%%%%%%%%%%%%%%%%%%%%%%%%%%%%%%%%%%%%%%%%%%%%%%%%%%%%%%
%
%
In this section, we acquire the general solution of (\ref{lind_Int}) formally. First to eliminate the commutator and anticommutators, 
we do the transformation
\begin{equation*}
\rho(t)=e^{(-i\Omega N-\gamma'/2)t}\tilde{\rho(t)}e^{(i\Omega^{*}N-\gamma'/2)t},
\ \Omega\equiv\omega-i\bar{\gamma},\ \bar{\gamma}\equiv \frac{\gamma+\gamma'}{2},\ N=\ad a.
\end{equation*}
Then (\ref{lind_Int}) becomes
\begin{equation}
\dot{\tilde{\rho}}=\gamma e^{-2\bar{\gamma}t}a\tilde{\rho}\ad+\gamma' e^{2\bar{\gamma}t}\ad\tilde{\rho}a. \label{LinKai_Der}
\end{equation}
To solve this equation we introduce the superoperators $\SK_1,\ \SK_2,\ \SK_3$ whose actions are 
\begin{equation}
\SK_1\rho = a\rho \ad,\ \SK_2\rho = \ad\rho a,\ \SK_3\rho=\left\{N+1/2,\rho\right\}.
\end{equation}
These superoperators have the algebraic structure of $\mathfrak{sl}(2)$.
\begin{equation*}
[\SK_1,\SK_2]=\SK_3,\ [\SK_3,\SK_1]=-2\SK_1,\ [\SK_3,\SK_2]=2\SK_2
\end{equation*}
Existence of this algebraic structure may allow us to expect to find a solution of the form
\begin{equation*}
\tilde{\rho}(t)=e^{f_2(t)\SK_2}e^{f_3(t)\SK_3}e^{f_1(t)\SK_1}\rho(0).
\end{equation*}
In terms of the superoperators, the time derivative of $\tilde{\rho}$ becomes
\begin{equation}
\dot{\tilde{\rho}}=\left\{\dot{f_1}e^{-2f_3}\SK_1+\left(\dot{f_2}-2f_2\dot{f_3}+\dot{f_1}e^{-2f_3}f_2^2\right)\SK_2
+\left(\dot{f_3}-\dot{f_1}e^{-f_3}f_2\right)\right\}
\tilde{\rho} \label{Dainyu_Der}
\end{equation}
Here we use the Hadamard lemma, i.e.\ for the elements $X, Y$ of Lie algebra $\mathfrak{g}$,
\begin{equation*}
e^YXe^{-Y}=X+[Y,X]+\frac{1}{2!}[Y,[Y,X]]+\cdots =e^{\mbox{ad}(Y)}X,
\end{equation*}
where $\mbox{ad}(Y)X\equiv[Y,X]$.
On the other hand, we can write down the right hand side of (\ref{LinKai_Der}) in terms of superoperators
\begin{equation}
\dot{\tilde{\rho}}=\gamma e^{-2\bar{\gamma}t}\SK_1\tilde{\rho}+\gamma' e^{2\bar{\gamma}t}\SK_2\tilde{\rho} \label{LinkaiT_Der}
\end{equation}
Comparing (\ref{Dainyu_Der}) with (\ref{LinkaiT_Der}), we notice that $f_j(t)\ (j=1, 2, 3)$ satisfy the simultaneous differential equation
\begin{eqnarray*}
&&\dot{f_1}e^{-2f_3}=\gamma e^{-2\bar{\gamma}t},\label{LinkaiT_df1}\\
&&\dot{f_2}-2f_2\dot{f_3}+\dot{f_1}e^{-2f_3}f_2^2=\gamma'e^{2\bar{\gamma}t},\\
&&\dot{f_3}-\dot{f_1}e^{-2f_3}f_2=0.\label{LinkaiT_df3}
\end{eqnarray*}
We solve this equation with the initial conditions $f_j(0)=0\ (j=1, 2, 3)$. First we can extract the differential equation of $f_2$
\begin{equation*}
\dot{f_2}-f_2^2\gamma e^{-2\bar{\gamma}t}-\gamma'e^{2\bar{\gamma}t}=0.
\end{equation*}
This is a Riccati equation. Once we obtain the solution of $f_2$, we can solve the other equations just by integration.
The solutions are
\begin{subequations}
\begin{eqnarray}
&&f_1(t)=\frac{1-e^{-\gmdf t}}{1-\tpn e^{-\gmdf t}},\\
&&f_2(t)=e^{2\bar{\gamma}t}\tpn\frac{1-e^{-\gmdf t}}{1-\tpn e^{-\gmdf t}},\\
&&f_3(t)=\gamma t-\ln\left(\frac{e^{\gmdf t}-\tpn}{\tmpn}\right).
\end{eqnarray}
\end{subequations}
Consequently, we obtain the formal solution of (\ref{lind_Int}) in terms of superoperators
\begin{equation}
\rho(t)=e^{(-i\Omega N-\gamma'/2)t}\{e^{f_2(t)\SK_2}e^{f_3(t)\SK_3}e^{f_1(t)\SK_1}\rho(0)\}e^{(i\Omega^{*}N-\gamma'/2)t}.\label{sol_Der}
\end{equation}

By using the following relations
\begin{eqnarray*}
&&e^{(-i\Omega N-\gamma'/2)t}\{e^{f_2(t)\SK_2}\rho\}e^{(i\Omega^{*}N-\gamma'/2)t}=e^{-\gamma' t}e^{f_2(t)e^{-2\bar{\gamma}t}\SK_2}\{e^{-i\Omega Nt}\rho e^{i\Omega^*Nt}\}\\
&&e^{f_3(t)\SK_3}\rho=e^{f_3(t)(N+1/2)}\rho e^{f_3(t)(N+1/2)},
\end{eqnarray*}
(\ref{sol_Der}) is rewritten as
\begin{equation}
\begin{aligned}
&\rho(t)=f(x)e^{-\tpn xf(x)\SK_2}e^{\tpn \SK_2}\left\{\left(e^{-i\omega t}x^{\frac{1}{2}}f(x)\right)^N\left(e^{-xf(x)\SK_1}e^{\SK_1}\rho(0)\right)\left(e^{i\omega t}x^{\frac{1}{2}}f(x)\right)^N\right\}\\
&x(t)\equiv e^{-\gmdf t},\ f(x)\equiv\frac{\tmpn}{1-\tpn x}.
\end{aligned}
\end{equation}
Here, we describe $N$ as a matrix in the basis of eigenstates and introduce some formulae
\begin{eqnarray*}
&&N=\sum^\infty_{n=0}n\ket{n}\bra{n},\nonumber\\
&&\ket{0}\bra{0}\{e^{\SK_1}\rho\}\ket{0}\bra{0}=\ket{0}\sum^\infty_{n=0}\bra{n}\rho\ket{n}\bra{0}=\ket{0}\bra{0}\tr\rho,\\
&&e^{\tpn\SK_2}\ket{0}\bra{0}=\sum^\infty_{n=0}e^{n\ln\tpn}\ket{n}\bra{n}=\eq.
\end{eqnarray*} 
The superoperator $e^{\SK_1}$ brings about essentially the trace operation, 
and acting $e^{\SK_2}$ on $\ket{0}\bra{0}$ makes the equilibrium density matrix. Using above formulae,
\begin{eqnarray*}
\rho(t)&=&\sum^\infty_{m,n=0}\frac{1}{m!n!}e^{-i(m-n)\omega t}x^{\frac{m+n}{2}}f(x)^{m+n+1}e^{-\tpn xf(x)\SK_2}\\
&&\quad \times\left(\eqw{m}{n}\right)\ttr{a^m\left(e^{-xf(x)\SK_1}\rho(0)\right)(\ad)^n}\\
&=&\sum^\infty_{m,n,p,q=0}\frac{(-1)^{p+q}}{m!n!p!q!}\tp^q e^{-i(m-n)\omega t}x^{\frac{m+n}{2}+p+q}f(x)^{m+n+p+q+1}\\
&&\quad \times\eqw{m+p}{n+p}\ttr{a^{m+q}\rho(0)(\ad)^{n+q}}.
\end{eqnarray*}
Expanding $f(x)$ with the formula
\begin{equation*}
\left(1-\tpn x\right)^{-s}=\left\{\sum^\infty_{r=0}\left(\tpn x\right)^r\right\}^s=\sum^\infty_{r=0}\frac{(r+s-1)!}{r!(s-1)!}\left(\tpn x\right)^r,\ s\in\mathbb{N},
\end{equation*}
we obtain the formal solution 
\begin{equation}
\begin{aligned}
\rho(t)=&\sum^\infty_{m,n,p,q,r=0}\frac{(-1)^{p+q}(m+n+p+q+r)!}{(m+n+p+q)!m!n!p!q!r!}\tmp^{m+n+p+q+1}\tp^{q+r}\\
& \quad \times e^{-\left(\frac{m+n}{2}+p+q+r\right)\gmdf t}e^{-i(m-n) \omega t}\eqw{m+p}{n+p}\ttr{a^{m+q}\rho(0)(\ad)^{n+q}}.\label{sol2_Der}
\end{aligned}
\end{equation}

Here we notice that the time dependent parts in (\ref{sol2_Der}) are only
\begin{equation*}
\exp\left[\left\{-\left(\frac{m+n}{2}+p+q+r\right)\gmdf-i(m-n)\omega \right\}t\right].
\end{equation*} 
Assembling the combinations $(m,n,p,q,r)$ which have the same exponent, we can write the solution in the form of (\ref{sol_Int}). 
For example,
\begin{xalignat*}{3}
(0,0,0,0,0)&: & &\lambda_{0,0}=0,                          & &\Pi_{0,0}\rho=\tmp \eq\tr\rho,\\
(1,0,0,0,0)&: & &\lambda_{1,1}=-\frac{\gmdfn}{2}-i\omega,  & &\Pi_{1,1}\rho=\tmp^2\ad \eq \ttr{a\rho},\\
(0,1,0,0,0)&: & &\lambda_{1,-1}=-\frac{\gmdfn}{2}+i\omega, & &\Pi_{1,-1}\rho=\tmp^2\eq a\ttr{\rho \ad},\\
(2,0,0,0,0)&: & &\lambda_{2,2}=-\gmdf-2i\omega,           & &\Pi_{2,2}\rho=\frac{1}{2}\tmp^3(\ad)^2\eq\ttr{a^2\rho},\\
(0,2,0,0,0)&: & &\lambda_{2,-2}=-\gmdf+2i\omega,          & &\Pi_{2,-2}\rho=\frac{1}{2}\tmp^3e^{-\beta\omega \ad a}a^2\ttr{\rho(\ad)^2},\\
%\end{xalignat*}
%\begin{xalignat*}{3}
(1,1,0,0,0)&: & &\lambda_{2,0}=-\gmdf,                    & &\Pi_{2,0}\rho=\tmp^3 \ad \eq a\ttr{a\rho \ad}\\
(0,0,1,0,0)&: & &                                         & &\hspace{1cm}-\tmp^2\tpn\ad\eq a\tr\rho\\
(0,0,0,1,0)&: & &                                         & &\hspace{1cm}-\tmp^2\eq\ttr{a\rho\ad}\\
(0,0,0,0,1)&: & &                                         & &\hspace{1cm}+\tmp\tpn\eq\tr\rho.
\end{xalignat*}
Executing this program, we presume that the eigenvalues are characterized as
\begin{equation}
\lambda_{j,k}=-\frac{j}{2}\gmdf-ik\omega,\ j\in\zeroN,\ k=-j, -j+2, \cdots, j-2, j,
\end{equation}
which agree with those in Ref.\cite{Tay1},
and the corresponding eigenprojections are, for $k\geq 0$ ($l\equiv(j-k)/2$),
\begin{subequations}
\begin{eqnarray}
\begin{aligned}
\Pi_{j,k}\rho=&\sum^l_{m,n=0}\frac{(-1)^{m+n}(k+l)!l!}{(k+m)!(k+n)!(l-m)!(l-n)!m!n!}\tmp^{k+m+n+1}\tp^{l-n}\\
&\quad \times\eqw{k+m}{m}\ttr{a^{k+n}\rho(\ad)^n},
\end{aligned}\\
\begin{aligned}
\Pi_{j,-k}\rho=&\left(\Pi_{j,k}\rho\right)^\dag\\
=&\sum^l_{m,n=0}\frac{(-1)^{m+n}(k+l)!l!}{(k+m)!(k+n)!(l-m)!(l-n)!m!n!}\tmp^{k+m+n+1}\tp^{l-n}\\
&\quad \times\eqw{m}{k+m}\ttr{a^{n}\rho(\ad)^{k+n}}.
\end{aligned}
\end{eqnarray}
\end{subequations}
Eventually we have obtained all the presumed eigenvalues and eigenprojections.
%
%
%%%%%%%%%%%%%%%%%%%%%%%%%%%%%%%%%%%%%%%%%%%%%%%%%%%%%%%   
\section{Property of the eigenprojections}\label{SEC_PrEi}
%%%\label{***_PrEi}
%%%%%%%%%%%%%%%%%%%%%%%%%%%%%%%%%%%%%%%%%%%%%%%%%%%%%%%
%
%
In this section we will show that the eigenvalues and eigenprojections which we obtained in the last section certainly satisfy

\begin{eqnarray*}
&&\SL\Pi_{j,k}=\Pi_{j,k}\SL=\lambda_{j,k}\Pi_{j,k},\\
&&\Pi_{j,k}\Pi_{j',k'}=\delta_{j,j'}\delta_{k,k'}\Pi_{j,k}\ (\mbox{orthogonality and idempotence}),\\
&&\sum_{j,k}\Pi_{j,k}=\mathbf{1}\ (\mbox{completeness}).
\end{eqnarray*}

First, we notice that the eigenprojections can be divided into the operator part and the trace part.
We expect that the eigenprojections have the structure of biorthogonal system. To explain this idea, we introduce the bra-ket notation for the vector
space of operators. We describe a Hilbert-Schmidt operator $A\in\hlsm$ as a ket $\kket{A}$. Since $\hlsm$ is a Hilbert space, we can also treat $B^\dag\in\hlsm$ as a bra $\bbra{B}$.
We define the Hilbert-Schmidt inner product between bras and kets $\bbrakket{B|A}\equiv\tr B^\dag A$. When we restrict the ket space to $\trcl$, the bra space
becomes $\bdop$. The pairing is defined similarly as $\bbrakket{B|A}\equiv\tr B^\dag A,\ A\in\trcl,\ B\in\bdop$. Let $\SO$ be an superoperator on the ket space.
The adjoint superoperator $\SO^\dag$ is the operator defined on the bra space whose action is defined as, $\bbrakket{\SO^\dag B|A}\equiv\bbra{B}\SO\kket{A}$.

Now we get back to the discussion of the eigenprojection. For $k\geq 0$, we define
\begin{subequations}
\begin{eqnarray}
&&\kket{j,k}\equiv \sqrt{C_{k,l}}\sum^l_{m=0}A_{k,l,m}\eqw{k+m}{m},\ l\equiv\frac{j-k}{2},\\
&&\bbra{j,k}\equiv \sqrt{C_{k,l}}\sum^l_{n=0}B_{k,l,n}(\ad)^n a^{k+n},\\
&&A_{k,l,m}\equiv \frac{(-1)^m}{(k+m)!(l-m)!m!}\tmp^m,\\
&&B_{k,l,n}\equiv \frac{(-1)^n}{(k+n)!(l-n)!n!}\tmp^n\tp^{-n},\\
&&C_{k,l}\equiv (k+l)!l!\tmp^{k+1}\tp^l.
\end{eqnarray}
\end{subequations}
$\kket{j,k}$ is shown to be the right eigenvector of the Liouvillian corresponding to the eigenvalue $\lambda_{j,k}$, 
and $\bbra{j,k}$ to be the left (generalized) eigenvector. Then the presumed eigenprojection can be written as
\begin{equation}
\Pi_{j,k}=\kket{j,k}\bbra{j,k}.
\end{equation}
For $k<0$ these eigenvectors are defined similarly. Consequently we obtain the spectral resolution of the Liouvillian as
\begin{equation}
\SL=\sum^\infty_{j=0}\sum^j_{l=0}\lambda_{j,j-2l}\kket{j,j-2l}\bbra{j,j-2l}. \label{Spec_PrEi}
\end{equation}

It is easy to see that $\eqw{m}{n},\ m,n\in\zeroN$ is a bounded and trace class operator,
whose trace is
\begin{equation}
\ttr{a^m(\ad)^n\eq}=\delta_{m,n}\frac{m!}{(1-e^{-\beta\omega})^{m+1}}.
\end{equation}
This formula is derived easily by induction. Therefore the right eigenvectors belong to the space of trace class operators $\trcl$.
This is desirable because axiomatically the space of the density matrix is in $\trcl$.
However the left eigenvector is obviously unbounded and not in $\trcl$. This problem is treated in section \ref{SEC_Math}.

Now what we should prove is
\begin{subequations}
\begin{eqnarray}
&&\SL\kket{j,k}=\lambda_{j,k}\kket{j,k},\ \bbra{j,k}\SL=\lambda_{j,k}\bbra{j,k},\label{Eivt_PrEi}\\
&&\bbrakket{j,k|j',k'}=\delta_{j,j'}\delta_{k,k'}\ (\mbox{orthonormality}),\label{NorOrt_PrEi}\\
&&\sum_{j,k}\kket{j,k}\bbra{j,k}=\mathbf{1}\ (\mbox{completeness}).\label{Comp_PrEi}
\end{eqnarray}
\end{subequations}

Let $\SL^\dag$ be the adjoint operator of the Liouvillian with respect to the Hilbert-Schmidt inner product,
\begin{equation*}
\SL^\dag\rho=i[H,\rho]+\gamma\left(\ad\rho a-\frac{1}{2}\{\ad a,\rho\}\right)+\gamma'\left(a\rho \ad-\frac{1}{2}\{a\ad,\rho\}\right).
\end{equation*}
For $k, m, n \in\zeroN$
\begin{eqnarray*}
&&\begin{aligned}
\SL\{\eqw{k+m}{m}\}&=\left\{-ik\omega-\frac{1}{2}(k+2m)\gmdf\right\}\eqw{k+m}{m}\\
&\quad+\gamma(k+m)m\eqw{k+m-1}{m-1}\\
\SL\{\eqw{m}{k+m}\}&=\left\{ik\omega-\frac{1}{2}(k+2m)\gmdf\right\}\eqw{m}{k+m}\\
&\quad+\gamma(k+m)m\eqw{m-1}{k+m-1}
\end{aligned}\\
&&\begin{aligned}
&\SL^\dag\{(\ad)^{k+n}a^n\}=\left\{ik\omega-\frac{1}{2}(k+2n)\gmdf\right\}(\ad)^{k+n}a^n+\gamma'(k+n)n(\ad)^{k+n-1}a^{n-1}\\
&\SL^\dag\{(\ad)^{n}a^{k+n}\}=\left\{-ik\omega-\frac{1}{2}(k+2n)\gmdf\right\}(\ad)^{n}a^{k+n}+\gamma'(k+n)n(\ad)^{n-1}a^{k+n-1}.
\end{aligned}
\end{eqnarray*}
Using above results, we can show (\ref{Eivt_PrEi}).

Next we confirm the orthonormality. Althogh the orthogonality is trivial from the property of the eigenvectors,
we check this by explicit calculation.
\begin{equation}
\bbrakket{j',k'|j,k}=\sqrt{C_{k,l}C_{k',l'}}\sum^l_{m=0}\sum^{l'}_{n=0}A_{k,l,m}B_{k',l',n}\ttr{a^m(\ad)^n a^{k'+n}(\ad)^{k+m}e^{-\beta\omega \ad a}}\label{inpro_PrEi}
\end{equation}
Apparently the trace vanishes unless $k=k'$. For $h\in\zeroN$, 
\begin{eqnarray*}
\bra{h}a^m(\ad)^n a^{k+n}\eql{k+m}\ket{h}&=&\{h+m-(n-1)\}\{h+m-(n-2)\}\cdots(h+m)\\
&&\quad\times(h+1)(h+2)\cdots(h+k+m)e^{-h\beta\omega}
\end{eqnarray*}
Setting $\xi\equiv e^{-\beta\omega}$,
\begin{equation*}
(h+\alpha)\xi^h=\xi^{-(\alpha-1)}\dif{\xi}\xi^\alpha\cdot \xi^h
\end{equation*}
Using above equation,
\begin{eqnarray*}
\bra{h}a^m(\ad)^n a^{k+n}\eql{k+m}\ket{h}
&=&\xi^{-(m-n)}\dif{\xi}\xi^{m-n+1}\cdot \xi^{-(m-n+1)}\dif{\xi}\xi^{m-n+2}\cdots \xi^{-(m-1)}\dif{\xi}\xi^j\\
&&\quad\times \xi^0\dif{\xi}\xi\cdot \xi^{-1}\dif{\xi}\xi^2\cdots \xi^{-(k+m-1)}\dif{\xi}\xi^{k+m} \cdot \xi^h\\
&=&\xi^{n-m}\left(\dif{\xi}\right)^n\xi^m\left(\dif{\xi}\right)^{k+m}\xi^{k+m+h}\\
\ttr{a^m(\ad)^n a^{k+n}\eql{k+m}}&=&\xi^{n-m}\left(\dif{\xi}\right)^n\xi^m\left(\dif{\xi}\right)^{k+m}\frac{\xi^{k+m}}{1-\xi}\\
&=&(k+m)!\xi^{n-m}\left(\dif{\xi}\right)^n\frac{\xi^m}{(1-\xi)^{k+m+1}}\\
&=&\frac{m!n!\xi^n}{(1-\xi)^{k+m+n+1}}\sum^{\min(m,n)}_{\alpha=0}\frac{(k+m+n-\alpha)!}{(m-\alpha)!(n-\alpha)!\alpha!}\left(\frac{1-\xi}{\xi}\right)^\alpha
\end{eqnarray*}
Substituting this trace to (\ref{inpro_PrEi})
\begin{eqnarray*}
\bbrakket{j',k'|j,k}&=&\delta_{k,k'}\frac{\sqrt{C_{k,l}C_{k,l'}}}{(1-\xi)^{k+1}}\sum^l_{m=0}\sum^{l'}_{n=0}\frac{(-1)^{m+n}}{(k+m)!(k+n)!(l-m)!(l'-n)!}\\
&&\quad\times\sum^{\min(m,n)}_{\alpha=0}\frac{(k+m+n-\alpha)!}{(m-\alpha)!(n-\alpha)!\alpha!}\left(\frac{1-\xi}{\xi}\right)^\alpha\\
&=&\delta_{k,k'}\frac{\sqrt{C_{k,l}C_{k,l'}}}{(1-\xi)^{k+1}\xi^l}\sum^{l'}_{n=0}\frac{(-1)^n}{(k+n)!(l'-n)!}\\
&&\quad\times\sum^l_{m=0}\sum^{\min(m,n)}_{\alpha=0}\frac{(-1)^m(k+m+n-\alpha)!}{(k+m)!(l-m)!(m-\alpha)!(n-\alpha)!\alpha!}(1-\xi)^\alpha \xi^{l-\alpha}.
\end{eqnarray*}
If we analyze the above sum for $l\geq l'$, it suffices. We claim that
\begin{equation}
\sum^l_{m=0}\sum^{\min(m,n)}_{\alpha=0}\frac{(-1)^m(k+m+n-\alpha)!}{(k+m)!(l-m)!(m-\alpha)!(n-\alpha)!\alpha!}(1-\xi)^\alpha \xi^{l-a}=\delta_{l,n}\frac{(-1)^l}{l!}.\label{Claim1_PrEi}
\end{equation}
This immediately leads to the orthonormality (\ref{NorOrt_PrEi}).

Now we prove (\ref{Claim1_PrEi}). Expanding $(1-\xi)^\alpha$ and changing the order of the summation, the left hand side of (\ref{Claim1_PrEi}) becomes
\begin{eqnarray*}
&&\sum^l_{m=0}\sum^{\min(m,n)}_{\alpha=0}\sum^\alpha_{\beta=0}\frac{(-1)^{m+\alpha+\beta}(k+m+n-\alpha)!}{(k+m)!(l-m)!(m-\alpha)!(n-\alpha)!(\alpha-\beta)!\beta!}\xi^{l-\beta}\\
&&=\sum^n_{\beta=0}\frac{(-1)^\beta}{\beta!}\xi^{l-\beta}\sum^{n}_{\alpha=\beta}\frac{(-1)^\alpha}{(n-\alpha)!(\alpha-\beta)!}\sum^l_{m=\alpha}\frac{(-1)^m(k+m+n-\alpha)!}{(k+m)!(l-m)!(m-\alpha)!}.
\end{eqnarray*}
Here we introduce a formula. For $p,q,r\in\zeroN,\ p\geq q\geq r$,
\begin{equation}
\sum^q_{s=0}(-1)^s\frac{(p+s-r)!}{(p+s-q)!(q-s)!s!}=(-1)^q\delta_{r,0},
\end{equation}
which is proved in Appendix \ref{APP_Sum}.
We apply this formula to the summation about $m$ by setting $p\equiv k+l,\ q\equiv l-\alpha,\ r\equiv l-n,\ s\equiv m-\alpha$.
Futhermore we introduce another formula. For $p,r\in\zeroN,\ p\geq r$
\begin{equation}
\sum^p_{q=r}(-1)^q\frac{p!}{(p-q)!(q-r)!}=(-1)^r r!\delta_{p,r}. \label{sum_Calc}
\end{equation}
Applying this formula to the summation about $\alpha$, we immediately obtain the right hand side of (\ref{Claim1_PrEi}) and finish the proof of
the orthonormality. The formula (\ref{sum_Calc}) is heavily used in section \ref{SEC_Calc}.

Finally we prove the completeness. If we ignore the left eigenvector problem temporarily, it is sufficient to prove that 
$\Span{\{\kket{j,k}\}_{j,k}}$ is dense in $\trcl$ and use the logic in appendix B. We see that
\begin{equation*}
\Span{\{\kket{j,k}\}_{j,k}}=\Span{\{\eqw{m}{n}\}^\infty_{m,n=0}}\equiv\mathfrak{S}
\end{equation*}
by the recombination of the basis. We will construct $\{\ket{m}\bra{n}\}^\infty_{m,n=0}$, which is the C.O.N.S of the Hilbert-Schmidt space $\hlsm$, 
as the limit of the point sequences on $\mathfrak{S}$. We use the next formulae \cite{Klau1}
\begin{eqnarray*}
&&:e^{\tau \ad a}:=e^{\log(1+\tau)\ad a}\\
&&:e^{-\ad a}:=\ket{0}\bra{0},
\end{eqnarray*}
where $:A:$ represents the normal ordering of the operator $A$. Setting $\tau\equiv e^{-\beta\omega}-1$,
\begin{eqnarray*}
&&\eq=:e^{\tau \ad a}:=\sum^\infty_{p=0}\frac{\tau^p}{p!}(\ad)^p a^p \\
&&:e^{-\ad a}:=\sum^\infty_{p=0}\frac{(-1)^p}{p!}(\ad)^p a^p=\ket{0}\bra{0}.
\end{eqnarray*}
By acting $\ad, a$ on the above equations,
\begin{eqnarray*}
&&\eqw{q+m}{q+n}=(\ad)^q\sum^\infty_{p=0}\frac{\tau^p}{p!}(\ad)^{p+m}a^{p+n}a^q\\
&&\ket{m}\bra{n}=\frac{1}{\sqrt{m!n!}}\sum^\infty_{p=0}\frac{(-1)^p}{p!}(\ad)^{p+m}a^{p+n}.
\end{eqnarray*}
Then we can obtain $\ket{m}\bra{n}$ from $\{\eqw{m}{n}\}^\infty_{m,n=0}$ as the following way,
\begin{equation*}
\ket{m}\bra{n}=\frac{1}{\sqrt{m!n!}}\sum^\infty_{q=0}\alpha_q\eqw{q+m}{q+n},\ \alpha_q=\frac{(-1)^q}{q!}-\sum^q_{p=1}\frac{\tau^p}{p!}\alpha_{q-p},\ \alpha_0=1.
\end{equation*}
We have constructed the C.O.N.S. of $\hlsm$ as the limit of the point sequences on $\mathfrak{S}$.
Therefore $\Span{\{\kket{j,k}\}_{j,k}}$ is dense in $\hlsm$ and dense in $\trcl$. The proof of the completeness taking the left eigenvector problem into account
will be done in section \ref{SEC_Math}.
%
%
%%%%%%%%%%%%%%%%%%%%%%%%%%%%%%%%%%%%%%%%%%%%%%%%%%%%%%%
\section{Rigged Hilbert space of the operators}\label{SEC_Math}
%%%\label{***_Math}
%%%%%%%%%%%%%%%%%%%%%%%%%%%%%%%%%%%%%%%%%%%%%%%%%%%%%%%
%
%
In this section we discuss the space which the eigenvectors of the Liouvillian belong to more rigorously.
The right eigenvectors are trace class operators, but the left eigenvectors are unbounded operators and not in $\trcl$.
To treat this problem we need the concept of the rigged Hilbert space. The Hilbert space which we deal in is the Hilbert-Schmidt space $\hlsm$. $\trcl$ is a Banach space with respect to the trace norm 
and dense subset of $\hlsm$. Therefore
\begin{equation*}
\trcl \subset \hlsm \subset \trcl'=\bdop
\end{equation*}
form a rigged Hilbert space. However the dual space is the space of bounded operators.

We need to think about more restricted space
\begin{equation}
\rsdm\equiv\{\rho\in\hlsm|a^m\rho(\ad)^n\in\trcl,\ \forall m,n\in\zeroN\}.
\end{equation}
with the seminorms
\begin{equation*}
\|A\|_{m,n}=\|a^mA(\ad)^n\|_1,\ m,n\in\zeroN.
\end{equation*}
This space becomes a Fr\'echet space.
For example $\eqw{m}{n}$ and $\ket{m}\bra{n}$ belong to $\rsdm$. The right eigenvectors are in $\rsdm$. 
$\rsdm$ includes the C.O.N.S of $\hlsm$, thus $\rsdm$ is dense in $\hlsm$ with respect to the topology of $\hlsm$. If we think about the triplet
\begin{equation}
\rsdm \subset \hlsm \subset \rsdm',
\end{equation}
$\rsdm'$ includes the operators like $(\ad)^m a^n$, which appear in the left eigenvectors. The left eigenvectors are in $\rsdm'$.

Now we consider the spectral resolution of the Liouvillian using the logic in appendix \ref{APP_Bio}. $\{\kket{j,k}\}_{j,k}$ is the set of the eigenvectors of the Liouvillian and
a complete system of $\rsdm$. From (\ref{NorOrt_PrEi}), $\{\kket{j,k},\bbra{j,k}\}_{j,k}$ is a biorthogonal system. 
Therefore $\{\bbra{j,k}\}_{j,k}$ is the set of generalized eigenvectors and complete in generalized meaning. Consequently the spectral resolution
(\ref{Spec_PrEi}) is justified.

Axiomatically, a density matrix is a normalized, positive semi-definite trace class operator, but above discussion imply that we should restrict the class of 
density matrix more strictly if we treat the unbounded operators as observables. For example,
\begin{equation*}
\frac{6}{\pi^2}\sum^\infty_{n=1}\frac{1}{n^2}\ket{n}\bra{n}
\end{equation*}
is a density matrix, but the energy expectation value of this density matrix is divergent. Therefore this restriction is physically reasonable.
%
%
%%%%%%%%%%%%%%%%%%%%%%%%%%%%%%%%%%%%%%%%%%%%%%%%%%%%%%%
\section{Time behavior of the solution}\label{SEC_Prop}
%%%%%%%%%%%%%%%%%%%%%%%%%%%%%%%%%%%%%%%%%%%%%%%%%%%%%%%
%
%
Eventually we have obtained the solution of the Lindblad master equation for an harmonic oscillator (\ref{lind_Int}).
\begin{equation}
\rho(t)=\sum^\infty_{j=0}\sum^{j}_{l=0}e^{\lambda_{j,j-2l}}\Pi_{j,j-2l}\rho(0)=
\sum^\infty_{j=0}\sum^{j}_{l=0}e^{-\frac{j}{2}\gmdf t}e^{-i(j-2l)\omega t}\Pi_{j,j-2l}\rho(0).
\end{equation}
This solution shows the time evolution of each eigenprojection explicitly. $e^{-\frac{j}{2}(\gamma-\gamma')t}$ has the effect of 
dissipation and $e^{-ik\omega t},\ k=j-2l$ represents the phase rotation, whose role become clear in section \ref{SEC_Calc}. Thus the eigenvalues
and the eigenprojections are classified according to the dissipation quantum number $j$ and the phase rotation quantum number $k$.
The only eigenprojection which belongs to $j=0$ and does not decay is
\begin{equation}
\Pi_{0,0}\rho(0)=\tmp \eq\tr\rho(0)=\left(1-e^{-\beta\omega}\right) \eq.
\end{equation}
This is nothing but the density matrix of the equilibrium state $\rho_{eq}$. Because other eigenprojections decay, the state converges to the equilibrium state 
as $t\to\infty$.

Next we confirm that this solution is appropriate for the time evolution of a density matrix. 
First we can show that only $\Pi_{0,0}\rho$ have nonzero traces. Apparently only $k=0$ eigenprojections has nonzero trace, therefore we study the 
case $j=2l, k=0, l\in\zeroN$.
\begin{equation*}
\bbrakket{\mathbf{1}|j,0}=\sqrt{C_{0,l}}\sum^l_{m=0}A_{0,l,m}\ttr{(\ad)^{m}e^{-\beta\omega \ad a}a^m}=\frac{\sqrt{C_{0,l}}}{l!\tmp}\sum^l_{m=0}(-1)^m\frac{l!}{m!(l-m)!}=\delta_{l,0}.
\end{equation*}
This leads to the trace preserving property $\tr\rho(t)=\tr\rho(0)\ (\forall t\in\mathbb{R}_{\geq 0})$, and the conservation of probability 
$\tr\dot{\rho}=0$. Self-adjointness of this solution is obvious. It is difficult to prove the positive-semidefiniteness by the explicit calculation.
However the Liouvillian is a generator of complete positive dynamical semigroup, and the solution of Lindblad equation must preserve the positivity.
%
%
%%%%%%%%%%%%%%%%%%%%%%%%%%%%%%%%%%%%%%%%%%%%%%%%%%%%%%%
\section{Calculation of some physical quantities}\label{SEC_Calc}
%%%\label{***_Calc}
%%%%%%%%%%%%%%%%%%%%%%%%%%%%%%%%%%%%%%%%%%%%%%%%%%%%%%%
%
%
In this section we calculate the expectation values and the variances of some physical quantities.
The formula (\ref{sum_Calc}), which is introduced in section \ref{SEC_PrEi}, plays an important role in the following calculation.
First we calculate the expectation value of energy $\braket{E}_t=\ttr{\rho(t)H}$.
\begin{eqnarray*}
\braket{E}_t&=&\omega\sum^\infty_{j=0}\sum^j_{l=0}e^{-\frac{j}{2}\gmdf t}e^{-i(j-2l)\omega t}
\bbrakket{\ad a|j,j-2l}\bbrakket{j,j-2l|\rho(0)}\\
&=&\omega\sum^\infty_{l=0}e^{-l\gmdf t}\bbrakket{\ad a|2l,0}\bbrakket{2l,0|\rho(0)}.\label{Et_Calc}
\end{eqnarray*}
Here we calculate $\bbrakket{\ad a|2l,0}$.
\begin{eqnarray*}
\bbrakket{\ad a|2l,0}&=&\sqrt{C_{0,l}}\sum^l_{m=0}A_{0,l,m}\tr[\ad a\eqw{m}{m}]\nonumber\\
&=&\frac{\sqrt{C_{0,l}}}{l!}\tmp^{-2}\sum^l_{m=0}(-1)^m\frac{l!}{(l-m!)m!}(m+\tpn)\nonumber\\
&=&\frac{\sqrt{C_{0,l}}}{l!}\tmp^{-2}\left(\delta_{l,0}\tpn-\delta_{l,1}\right).\label{aa2k_Calc}
\end{eqnarray*}
Thus the expectation value of energy is
\begin{eqnarray}
\braket{E}_t&=&\omega\sum^\infty_{l=0}e^{-k\gmdf t}\frac{\sqrt{C_{0,l}}}{l!}\tmp^{-2}\left(\delta_{l,0}\tpn-\delta_{l,1}\right)\bbrakket{2l,0|\rho(0)}\nonumber\\
&=&\omega\frac{\tpn}{\tmpn}-e^{-\gmdf t}\left[\omega\frac{\tpn}{\tmpn}-\omega\ttr{\rho(0) \ad a}\right]\nonumber\\
&=&\braket{E}_{eq}-e^{-\gmdf t}\left(\braket{E}_{eq}-\braket{E}_0\right),
\end{eqnarray}
where $\braket{\cdot}_{eq}$ means the expectation value at the equilibrium state, that is $\ttr{\rho_{eq}H}$. The convergence
$\braket{E}_t\to\braket{E}_{eq},\ t\to\infty$ is consistent with the convergence of the state. 
This result agrees with that in Ref. \cite{Breu1}.
We notice that because of the formula (\ref{sum_Calc}) only the eigenprojections $(j,k)=(0,0), (2,0)$ contribute to the calculation. 
Similarly the variance of energy is
\begin{eqnarray}
&&\begin{aligned}
\braket{E^2}_t&=\braket{E^2}_{eq}-e^{-\gmdf t}\left(\braket{E^2}_{eq}+2\braket{E}_{eq}^2-4\braket{E}_{eq}\braket{E}_0-\omega\braket{E}_0\right)\\
&\quad+e^{-2\gmdf t}\left(2\braket{E}_{eq}^2-4\braket{E}_{eq}\braket{E}_0-\omega\braket{E}_0+\braket{E^2}_0\right)
\end{aligned}\\
&&\begin{aligned}
(\Delta E_t)^2&=\braket{E^2}_t-\braket{E}_t^2\\
&=(\Delta E_{eq})^2-e^{-\gmdf t}\left(\braket{E}_{eq}^2-2\braket{E}_{eq}\braket{E}_0-\omega\braket{E}_0\right)\\
&\quad+e^{-2\gmdf t}\left(\braket{E^2}_{eq}-2\braket{E}_{eq}\braket{E}_0-\omega\braket{E}_0+(\Delta E_0)^2\right)
\end{aligned}
\end{eqnarray}
In this calculation the eigenprojections $(j,k)=(0,0), (2,0), (4,0)$ contribute to the second moment of energy. In general the eigenprojections
$(j,k)=(0,0), (2,0), \cdots, (2p,0)$ contribute to the $p$-th moment of energy. 
As expected, these results describe the convergence from the initial values to the equilibrium values. If we know the initial and equilibrium energy 
expectation value, we can calculate the time behavior of energy expectation value exactly.

Next we calculate the expectation values and variances of position and momentum. The position and momentum operators are described in terms of 
creation and annihilation operators.
\begin{equation*}
x=\frac{1}{\sqrt{2\omega}}(a+\ad),\ p=-i\sqrt{\frac{\omega}{2}}(a-\ad).
\end{equation*}
The expectation value of position is
\begin{eqnarray}
\braket{x}_t&=&\frac{1}{\sqrt{2\omega}}\ttr{(a+\ad)\rho(t)}\nonumber\\
&=&\frac{1}{\sqrt{2\omega}}\tmp^{-2}\sqrt{C_{1,0}}\left\{e^{\lambda_{1,1}t}\bbrakket{1,1|\rho(0)}+e^{\lambda_{1,-1}t}\bbrakket{1,-1|\rho(0)}\right\}\nonumber\\
&=&\frac{1}{\sqrt{2\omega}}e^{-\frac{1}{2}\gmdf t}\left[e^{-i\omega t}\ttr{a\rho(0)}+e^{i\omega t}\ttr{\ad\rho(0)}\right]\nonumber\\
&=&\sqrt{\frac{2}{\omega}}e^{-\frac{1}{2}\gmdf t}\mbox{Re}\left[e^{-i\omega t}\ttr{a\rho(0)}\right]
\end{eqnarray}
and that of momentum is
\begin{eqnarray}
\braket{p}_t&=&-i\sqrt{\frac{\omega}{2}}e^{-\frac{1}{2}\gmdf t}\left[e^{-i\omega t}\ttr{a\rho(0)}-e^{i\omega t}\ttr{\ad\rho(0)}\right]\nonumber\\
&=&\sqrt{2\omega}e^{-\frac{1}{2}\gmdf t}\mbox{Im}\left[e^{-i\omega t}\ttr{a\rho(0)}\right].
\end{eqnarray}
These results also describe the convergence from the initial values to the equilibrium values. The $(m,n)=(1,1), (1,-1)$ eigenprojections contribute to
the expectation values. Similarly 
\begin{eqnarray}
&&\begin{aligned}
\braket{x^2}_t&=\braket{x^2}_{eq}+\frac{1}{2\omega}e^{-\gmdf t}\left[e^{-2i\omega t}\ttr{a^2\rho(0)}+e^{2i\omega t}\ttr{(\ad)^2\rho(0)}\right.\\
&\quad\left.+\ttr{\ad a\rho(0)}+\ttr{a\ad\rho(0)}-2\omega\braket{x^2}_{eq}\right]\\
\end{aligned}\\
&&\begin{aligned}
\braket{p^2}_t&=\braket{p^2}_{eq}-\frac{\omega}{2}e^{-\gmdf t}\left[e^{-2i\omega t}\ttr{a^2\rho(0)}+e^{2i\omega t}\ttr{(\ad)^2\rho(0)}\right.\\
&\quad\left.-\ttr{\ad a\rho(0)}-\ttr{a\ad\rho(0)}+\frac{2}{\omega}\braket{p^2}_{eq}\right]\\
\end{aligned}\\
&&(\Delta x_{eq})^2=\braket{x^2}_{eq}=\frac{1}{2\omega}\frac{1+\tpn}{\tmpn},\ (\Delta p_{eq})^2=\braket{p^2}_{eq}=\frac{\omega}{2}\frac{1+\tpn}{\tmpn}.
\end{eqnarray}
For example, we consider the initial state
\begin{equation*}
\rho(0)=\frac{1}{2}\left(\ket{0}\bra{0}+\ket{1}\bra{1}+\ket{0}\bra{1}+\ket{1}\bra{0}\right).
\end{equation*}
The time behavior of the expectation values of position and momentum is
\begin{eqnarray*}
\braket{x}_t&=&\frac{1}{\sqrt{2\omega}}e^{-\frac{1}{2}\gmdf t}\cos\omega t\\
\braket{p}_t&=&-\sqrt{\frac{\omega}{2}}e^{-\frac{1}{2}\gmdf t}\sin\omega t.
\end{eqnarray*}
This is classically the motion of a damping oscillator. The time behavior of variance is
\begin{eqnarray*}
\braket{x^2}_t&=&\left(1-e^{-\gmdf t}\right)\braket{x^2}_{eq}+\frac{1}{\omega}e^{-\gmdf t}\\
\braket{p^2}_t&=&\left(1-e^{-\gmdf t}\right)\braket{p^2}_{eq}+\omega e^{-\gmdf t}\\
(\Delta x_t)^2&=&\left(1-e^{-\gmdf t}\right)(\Delta x_{eq})^2+\frac{1}{2\omega}e^{-\gmdf t}(1+\sin^2\omega t)\\
(\Delta p_t)^2&=&\left(1-e^{-\gmdf t}\right)(\Delta p_{eq})^2+\frac{1}{2\omega}e^{-\gmdf t}(1+\cos^2\omega t).
\end{eqnarray*}
%
%
%%%%%%%%%%%%%%%%%%%%%%%%%%%%%%%%%%%%%%%%%%%%%%%%%%%%%%%
\section{Construction of ladder operators}\label{SEC_Lad}
%%%\label{***_Lad}
%%%%%%%%%%%%%%%%%%%%%%%%%%%%%%%%%%%%%%%%%%%%%%%%%%%%%%%
%
%
In this section, we construct the ladder operators of the Liouvillian.
Let $L_+, L_-, R_+, R_-$ be the superoperators which have the action
\begin{equation*}
L_+\rho=\ad\rho,\ L_-\rho=a\rho,\ R_+\rho=\rho a,\ R_-\rho=\rho \ad.
\end{equation*}
These superoperators satisfy the following algebraic structure.
\begin{equation*}
[L_-,L_+]=[R_-,R_+]=1,\ \mbox{others}=0.
\end{equation*}
The Liouvillian can be written in terms of these superoperators.
\begin{equation*}
\SL=-i\omega(L_+L_- - R_+R_-)+\gamma L_-R_-+\gamma' L_+R_+-\frac{\gamma+\gamma'}{2}(L_+L_- + R_+R_-)-\gamma'.
\end{equation*}
The commutation relation between the Liouvillian and the superoperators is

\begin{eqnarray*}
&&[\SL,L_+]=-\left(i\omega+\frac{\gamma+\gamma'}{2}\right)L_{+} +\gamma R_{-}\\{}
&&[\SL,L_-]=\left(i\omega+\frac{\gamma+\gamma'}{2}\right)L_- -\gamma' R_+\\{}
&&[\SL,R_+]=\left(i\omega  -\frac{\gamma+\gamma'}{2}\right)R_+ +\gamma L_- \\{}
&&[\SL,R_-]=-\left(i\omega -\frac{\gamma+\gamma'}{2}\right)R_- -\gamma' L_+.
\end{eqnarray*}

We notice that $(\SL,L_+,R_-)$ and $(\SL,L_-,R_+)$ form closed algebras respectively, hence we set
\begin{eqnarray}
&&\SX_+\equiv L_+-R_-,\ \SX_-\equiv\gamma' R_+ -\gamma L_-\\
&&\SY_+\equiv R_+-L_-,\ \SY_-\equiv\gamma' L_+ -\gamma R_-.
\end{eqnarray}
The commutation relation between the Liouvillian and these new superoperators is
\begin{subequations}
\begin{eqnarray}
&&[\SL,\SX_+]=\left\{-\frac{1}{2}\gmdf-i\omega\right\}\SX_+\\{}
&&[\SL,\SX_-]=\left\{\frac{1}{2}\gmdf+i\omega\right\}\SX_-\\{}
&&[\SL,\SY_+]=\left\{-\frac{1}{2}\gmdf+i\omega\right\}\SY_+\\{}
&&[\SL,\SY_-]=\left\{\frac{1}{2}\gmdf-i\omega\right\}\SY_-\\{}
&&[\SX_-,\SX_+]=[\SY_-,\SY_+]=-\gmdf,\ \mbox{others}=0.
\end{eqnarray}
\end{subequations}
Therefore $\SX_+$ raises the dissipation quantum number $j$ and raise the phase rotation quantum number $k$,
$\SY_+$ raises $j$ and lowers $k$, $\SX_-$ lowers $j$ and lowers $k$, and $\SY_-$ lowers $j$ and raises $k$.
There are two independent pairs of radder operators $(\SX_+,\SX_-),\ (\SY_+,\SY_-)$. Now we determine the ground state. The ground state $\ktet{0}=\rho_0$ satisfies $\SX_-\ktet{0}=\SY_-\ktet{0}=0$ i.e.
\begin{equation*}
\gamma'\ad\rho_0-\gamma\rho_0\ad=0,\ \gamma'\rho_0 a-\gamma a\rho_0=0.
\end{equation*}
We want to find the solution of this equation in $\trcl$. Hence we expand $\rho_0$ on the basis $\{\ket{m}\bra{n}\}_{m,n}$.
Then this equation reduces to the equation of the expansion coefficient. We find
\begin{equation}
\ktet{0}=Ce^{-\beta\omega \ad a},\ C=\mbox{const}.
\end{equation}
Furthermore we obtain all eigenstate of the Liouvillian by acting $\SX_+,\SY_+$ on $\ktet{0}$.
\begin{equation}
\begin{aligned}
&\ktet{m,n}\equiv \SX_+^m\SY_+^n\ktet{0},\ m,n\in\zeroN \\
&\mathcal{L}\ktet{m,n}=\left\{-\frac{(m+n)}{2}\gmdf-i\omega(m-n)\right\}\ktet{m,n}\equiv\mu_{m,n}\ktet{m,n}.
\end{aligned}
\end{equation}
Comparing the eigenvalues and the eigenstates to the previous result
\begin{equation}
\mu_{m,n}=\lambda_{m+n,m-n},\ \ktet{m,n}\propto \kket{m+n,m-n}
\end{equation}
Next we construct dual vectors. The dual of the ground state $\btra{0}$ satisfies $\btra{0}\SX_+=\btra{0}\SY_+=0$, thus $\btra{0}\propto \mathbf{1}$.
Imposing the normalization condition $\btraktet{0|0}=1$,
\begin{equation}
\btra{0}=\frac{1}{C}\tmp\mathbf{1}.
\end{equation}
Therefore the dual vector of $\ktet{m,n}$ is
\begin{equation}
\btra{m,n}=\frac{1}{m!n!(\gamma'-\gamma)^{m+n}}\btra{0}\SX_-^m\SY_-^n.
\end{equation}
Now we have obtained the biorthogonal eigenbasis. We can write down the spectral resolution of $\SL$ and the solution of the Lindblad equation.
\begin{eqnarray}
&&\SL=\sum^\infty_{m,n=0}\mu_{m,n}\ktet{m,n}\btra{m,n}\\
&&\rho(t)=\sum^\infty_{m,n=0}e^{\mu_{m,n}t}\frac{1}{m!n!(\gamma'-\gamma)^{m+n}}\tmp\ttr{\SX_-^m\SY_-^n\rho(0)}\SX_+^m\SY_+^n\eq
\end{eqnarray}
This result is equivalent to the previous one.
%
%
%%%%%%%%%%%%%%%%%%%%%%%%%%%%%%%%%%%%%%%%%%%%%%%%%%%%%%%
%\section{Conclusion}
%%%\label{***_Con}
%%%%%%%%%%%%%%%%%%%%%%%%%%%%%%%%%%%%%%%%%%%%%%%%%%%%%%%
%
%

%%%%%%%%%%%%%%%%%%%%%%%%%%%%%%%%%%%%%%%%%%%%%%%%%%%%%%%
\begin{acknowledgments}
%%%%%%%%%%%%%%%%%%%%%%%%%%%%%%%%%%%%%%%%%%%%%%%%%%%%%%%
We would like to thank Professor Ichiro Ohba, Drs. Makoto Unoki (Waseda University), and Makoto Mine (Waseda University Honjo Senior High School) for 
insightful comments and suggestions. Especially D.H. thanks to Junichiro Noguchi, Toshifumi Noumi, Shingo Torii and Takuro Yoshimoto (University of Tokyo).
%%%%%%%%%%%%%%%%%%%%%%%%%%%%%%%%%%%%%%%%%%%%%%%%%%%%%%%
\end{acknowledgments}
%%%%%%%%%%%%%%%%%%%%%%%%%%%%%%%%%%%%%%%%%%%%%%%%%%%%%%%

\appendix
%
%
%%%%%%%%%%%%%%%%%%%%%%%%%%%%%%%%%%%%%%%%%%%%%%%%%%%%%%%
\section{Summary of trace class operators and Hilbert-Schmidt operators}%\label{APP_Sum}
%%%\label{***_Sum}
%%%%%%%%%%%%%%%%%%%%%%%%%%%%%%%%%%%%%%%%%%%%%%%%%%%%%%%
%
%
In this appendix we summarize the facts about the trace class operators and the Hilbert-Schmidt operators \cite{Reed1, Reed2} used in this work.
Let $\Hilb$ be a separable Hilbert space ,$\bdop$ be the set of the bounded operators on $\Hilb$, which is a $C^*$ algebra with the operator norm $\|A\|$,
and $\{\ket{n}\}_n$ be a C.O.N.S of $\Hilb$.
The set of the trace class operators $\trcl=\{A\in\bdop|\sum_n\bra{n}\sqrt{A^\dag A}\ket{n}<\infty\}$ is a $C^*$ algebra with the
trace norm $\|A\|_1\equiv \tr\sqrt{A^\dag A}\equiv \sum_n\bra{n}\sqrt{A^\dag A}\ket{n}$, and a two-sided $*$-ideal of $\bdop$.
The Hilbert-Schmidt space $\hlsm=\{A\in\bdop|A^\dag A\in\trcl\}$ is a Hilbert space with the Hilbert-Schmidt inner product 
$\bbrakket{A|B}\equiv\tr{A^\dag B}$, which is well-defined because $A,B\in\hlsm\Rightarrow AB\in\trcl, A^\dag\in\hlsm$,
a $C^*$ algebra with the norm $\|A\|_2\equiv\sqrt{\tr{A^\dag A}}$, and a two-sided $*$-ideal of $\bdop$. 
The relations between these three operator spaces and their norms are,
\begin{eqnarray}
&&\trcl\subset\hlsm\subset\bdop\\
&&\|A\|\leq\|A\|_2,\ A\in\hlsm,\ \|B\|_2\leq\|B\|_1,\ B\in\trcl\label{top_Not}.
\end{eqnarray}
From (\ref{top_Not}) the topology of $\trcl$ is finer than that of $\hlsm$. A C.O.N.S of $\hlsm$ is $\{\ket{m}\bra{n}\}_{m,n\in\zeroN}$. $\ket{m}\bra{n}$ is trace class, 
then $\Span{\{\ket{m}\bra{n}\}_{m,n}}\subset\trcl$ and $\trcl$ is dense in $\hlsm$.
%
%
%%%%%%%%%%%%%%%%%%%%%%%%%%%%%%%%%%%%%%%%%%%%%%%%%%%%%%%
\section{Biorthogonal system and rigged Hilbert space}\label{APP_Bio}
%%%\label{***_Bio}
%%%%%%%%%%%%%%%%%%%%%%%%%%%%%%%%%%%%%%%%%%%%%%%%%%%%%%%
%
%
First we explain the theory of the biorthogonal basis \cite{Mors1}. In general, for a non-Hermitian operator $A$ on a Hilbert space $\Hilb$, 
the eigenvalues of $A$ and those of $A^\dag$ are related by the complex conjugate. 
Let $\ket{m_R}\in \Hilb$ be the eigenvector corresponding to the $m$-th eigenvalue $\lambda_m$ of $A$, and $\ket{m_L}\in \Hilb$ be that of $A^\dag$.
Then 
\begin{equation}
A\ket{m_R}=\lambda_m\ket{m_R},\ A^\dag\ket{m_L}=\lambda_m^*\ket{m_L}.
\end{equation}
But generally $\ket{m_R}\neq\ket{m_L}$. If the eigenvalues are non-degenerate, $m\neq n \Rightarrow \braket{m_L|n_R}=0$.
If the eigenvectors are normalized, $\braket{m_L|n_R}=\delta_{m,n}$. 
We call the pair $\{\ket{m_R}, \bra{m_L}\}_m$ which satisfies $\braket{m_L|n_R}=\delta_{m,n}$ a biorthogonal system.
If $\{\ket{m_R}\}_m$ is a complete system, the completeness is expressed as $\sum_m\ket{m_R}\bra{m_L}=\mathbf{1}$.

Next we explain the concept of the rigged Hilbert space \cite{Gelf1}. 
Let $\mathscr{H}$ be a Hilbert space and $\Phi$ be a subset of $\mathscr{H}$ which satisfies the following properties.\\
1)$\Phi$ has a finer topology than that of $\mathscr{H}$.\\
2)$\Phi$ is complete with respect to the topology described at 1).\\
3)$\Phi$ is dense in $\mathscr{H}$ with respect to the topology of $\mathscr{H}$.\\
Then, the continuous anti-dual space of $\Phi$, denoted by $\Phi'$, is an extension of the Hilbert space. The triplet
\begin{equation}
\Phi\subset \mathscr{H} \subset \Phi'
\end{equation}
is called the the rigged Hilbert space or Gel'fand triplet. For example, let choose $L^2$ space as $\mathscr{H}$, the Schwartz space $\mathscr{S}$ as $\Phi$.
Then $\Phi'$ is the space of tempered distributions $\mathscr{S}'$.
\begin{equation}
\mathscr{S} \subset L^2 \subset \mathscr{S}'.
\end{equation}

Now we explain the idea of the generalized eigenvector. Let $\Phi\subset \mathscr{H} \subset \Phi'$ be a rigged Hilbert space, $A$ be a operator defined on 
$\Phi$, and $\ket{\xi}\in\Phi,\ \bra{\eta}\in\Phi'$. The adjoint operator $A^\dag$ is the operator defined on $\Phi'$ whose action is defined as
\begin{equation}
\braket{A^\dag \eta|\xi}:=\bra{\eta}A\ket{\xi}.
\end{equation}
If there exist $\bra{\eta_\lambda}\in\Phi'$ and $\lambda\in\mathbb{C}$ such that
\begin{equation}
\bra{A^\dag\eta_\lambda}=\lambda\bra{\eta_\lambda},
\end{equation}
that is, for any $\ket{\xi}\in\Phi$
\begin{equation}
\bra{\eta_\lambda}A\ket{\xi}=\lambda\braket{\eta_\lambda|\xi},
\end{equation}
then $\bra{\eta_\lambda}$ is called a generalized eigenvector of $A$ corresponding to the eigenvalue of $\lambda$. 
The system $\{\bra{\eta_m}\}_m$ is called complete when
\begin{equation}
\braket{\eta_m|\xi}=0\ \mbox{for all}\ m \Rightarrow \ket{\xi}=0.
\end{equation}

We apply the idea of the biorthogonal system to the rigged Hilbert space. Let $A$ be an operator defined on $\Phi$, $\{\ket{m}\}_m$ be the system of eigenvectors of $A$, that is
$A\ket{m}=\lambda_m\ket{m}$, and $\{\bra{n}\}_n$ be the system of generalized eigenvectors of $A$, that is, $\bra{A^\dag n}=\lambda_n\bra{n}$.
If the eigenvalues are non-degenerate, $m\neq n \Rightarrow \braket{m|n}=0$.
If the eigenvectors are normalized, $\braket{m|n}=\delta_{m,n}$.
We call the pair $\{\ket{m}, \bra{m}\}_m$ which satisfies $\braket{m|n}=\delta_{m,n}$ a biorthogonal system.
If $\{\ket{m}\}_m$ is a complete system, $\{\bra{m}\}_m$ is also complete in the generalized meaning, and the completeness is expressed as 
$\sum_m\ket{m}\bra{m}=\mathbf{1}$.\\

If $\{\ket{m}, \bra{m}\}_m$ is a biorthogonal system and $\{\ket{m}\}_m$ is the set of eigenvectors of $A$ and a complete system, 
$\{\bra{m}\}_m$ are the generalized eigenvectors of $A$. This is because any $\ket{\xi}\in\Phi$ can be expanded as $\ket{\xi}=\sum_mc_m\ket{m},\ c_m\in\mathbb{C}$, and
\begin{equation}
\braket{A^\dag n|\xi}=\bra{n}A\ket{\xi}=\sum_mc_m\lambda_m\braket{n|m}=c_n\lambda_n=\lambda_n\braket{n|\xi}.
\end{equation}
%
%
%%%%%%%%%%%%%%%%%%%%%%%%%%%%%%%%%%%%%%%%%%%%%%%%%%%%%%%
\section{Proof of the summation formula}\label{APP_Sum}
%%%\label{***_Sum}
%%%%%%%%%%%%%%%%%%%%%%%%%%%%%%%%%%%%%%%%%%%%%%%%%%%%%%%
%
%
In this appendix we prove the formula introduced in section \ref{SEC_PrEi}. This proof is due to Noguchi, Noumi, Yoshimoto. For $p,q,r\in\zeroN,\ p\geq q\geq r$,
\begin{equation}
\sum^q_{s=0}(-1)^s\frac{(p+s-r)!}{(p+s-q)!(q-s)!s!}=(-1)^q\delta_{r,0}.\label{prop_Sum}
\end{equation}

First, changing the variable $s\rightarrow q-s$, the formula (\ref{prop_Sum}) becomes
\begin{equation}
I(p,q,r)\equiv\sum^q_{s=0}(-1)^s\frac{(p+q-s-r)!}{(p-s)!(q-s)!s!}=\delta_{r,0}.\label{prop2_Sum}
\end{equation}
We can derive the recurrence formula
\begin{equation}
I(p,q,r-1)=(p+q-r+1)I(p,q,r)+I(p-1,q-1,r-1),
\end{equation}
and show that
\begin{equation}
I(p,q,q)=\delta_{q,0},\ p\geq q.
\end{equation}
First we prove (\ref{prop2_Sum}) in the case of $r>0$.
If $I(p,q,q-s)=0$ for some $s<q$,
\begin{equation}
I(p,q,q-(s+1))=(p+s+1)I(p,q,q-s)+I(p-1,q-1,(q-1)-s)=0.
\end{equation}
By mathematical induction, $I(p,q,r)=0$ for $p\geq q\geq r>0$.
In the case of $r=0$,
\begin{eqnarray}
I(p,q,0)&=&(p+q)I(p,q,1)+I(p-1,q-1,0)\nonumber\\
&=&I(p-1,q-1,0)=\cdots=I(p-q,0,0)=1.
\end{eqnarray}
%
%
%%%%%%%%%%%%%%%%%%%%%%%%%%%%%%%%%%%%%%%%%%%%%%%%%%%%%%%

%%%%%%%%%%%%%%%%%%%%%%%%%%%%%%%%%%%%%%%%%%%%%%%%%%%%%%%
%
%
\end{document}